\documentclass[aps,superscriptaddress,twocolumn,showpacs,prl]{revtex4}

\usepackage{graphicx}
\usepackage{bm}

\newcommand{\St}{\mbox{\it St}}

\newcommand{\Rey}{\mbox{\it R}}

\begin{document}

\title{Caustics and Intermittency in Turbulent Suspensions of  Heavy Particles}

\author{J.\ Bec} \affiliation{Universit\'{e} de Nice-Sophia Antipolis, CNRS,
  Observatoire de la C\^{o}te d'Azur, Laboratoire Cassiop\'{e}e,
  Bd.\ de l'Observatoire, 06300 Nice, France.}

\author{L.\ Biferale} \affiliation{Department of Physics and INFN,
  Universit\`a Tor Vergata, Via della Ricerca Scientifica 1, 00133
  Roma, Italy.}

\author{M.\ Cencini} \affiliation{INFM-CNR, SMC Dept.\ of Physics,
  Universit\`a ``La Sapienza'', P.zzle A.~Moro~2, and ISC-CNR, Via dei
  Taurini 19, 00185 Roma, Italy.}

\author{A. S.\ Lanotte} \affiliation{ISAC-CNR, Via Fosso del Cavaliere
  100, 00133 Roma, and INFN, Sezione di Lecce, 73100 Lecce Italy.}

\author{F.\ Toschi} \affiliation{Department of Physics and Department
  of Mathematics and Computer Science and J.M.\ Burgers Centre for
  Fluid Dynamics, Eindhoven University of Technology, 5600 MB
  Eindhoven, The Netherlands.} \affiliation{IAC-CNR, Viale del
  Policlinico 137, 00161 Roma, Italy.}

\begin{abstract}
  The statistics of velocity differences between very heavy inertial
  particles suspended in an incompressible turbulent flow is found to
  be extremely intermittent. When particles are separated by distances
  within the viscous subrange, the competition between quiet regular
  regions and multi-valued caustics leads to a quasi bi-fractal
  behavior of the particle velocity structure functions, with
  high-order moments bringing the statistical signature of
  caustics. Contrastingly, for particles separated by inertial-range
  distances, the velocity-difference statistics is characterized in
  terms of a local H\"{o}lder exponent, which is a function of the
  scale-dependent particle Stokes number only. Results are supported
  by high-resolution direct numerical simulations. It is argued that
  these findings might have implications in the early stage of rain
  droplets formation in warm clouds.
\end{abstract}
\date{\today}
\maketitle

\begin{figure}[b]
  \includegraphics[width=.42\columnwidth]{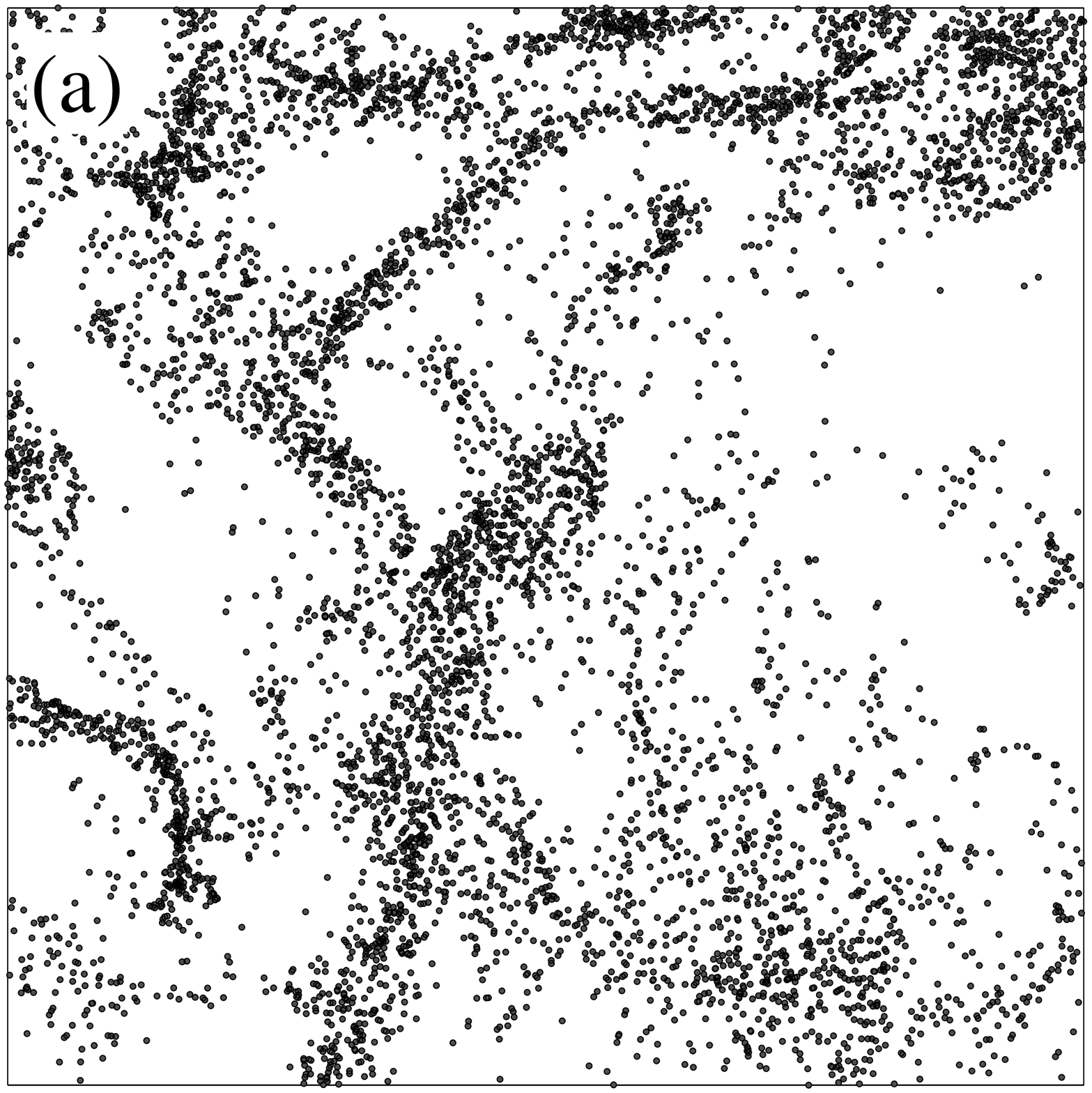}
  \includegraphics[width=.42\columnwidth]{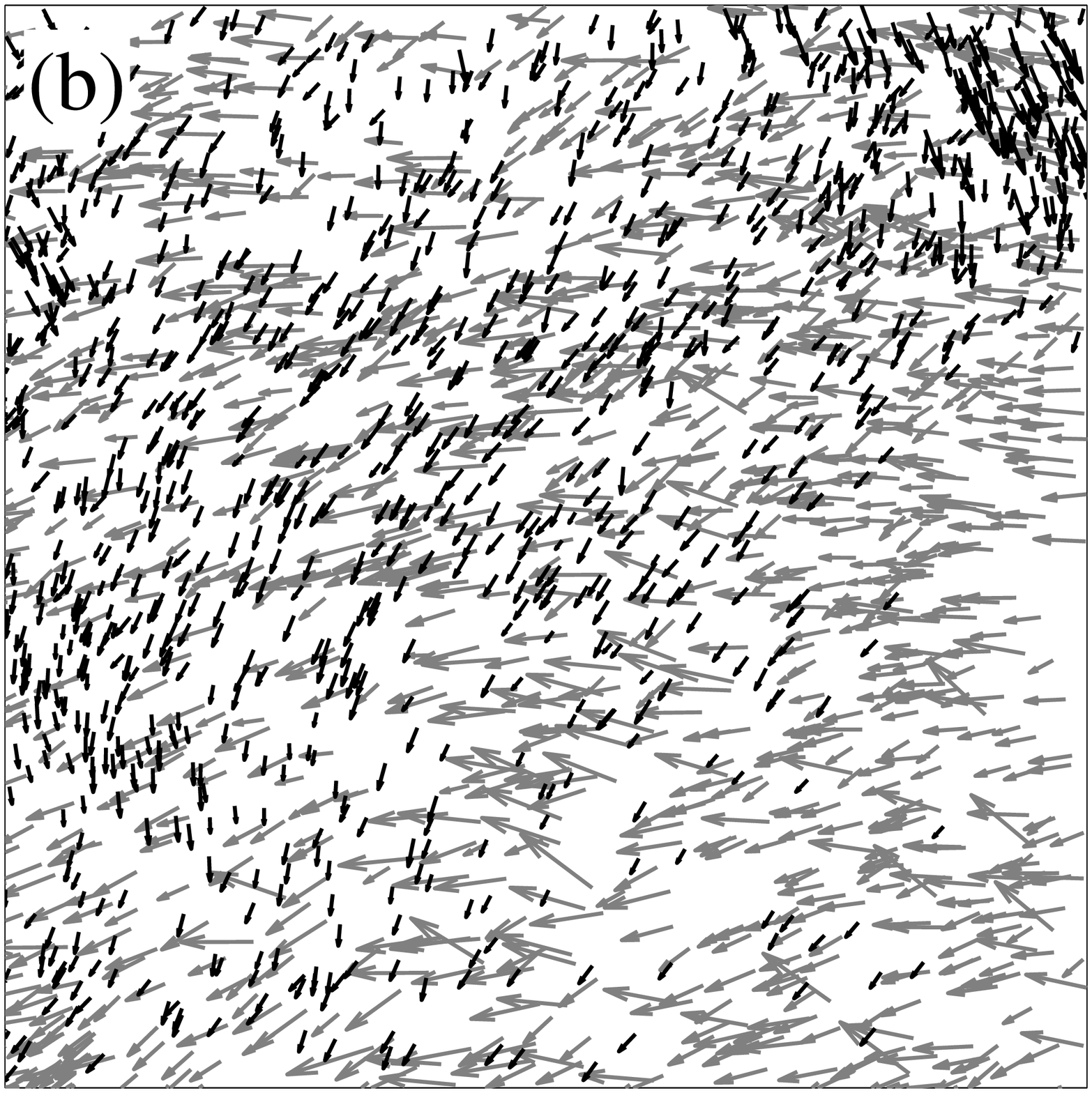}
\vspace{-12pt}
\caption{\label{fig:twomechanisms} (a) Snapshot of the position of
  particles for $\St=2$ in a slice of size $5\eta\!\times\! 100\eta\!
  \times\! 100\eta$ for $\Rey_\lambda\approx400$. (b) Particle
  velocity field in the same slice for a larger Stokes, $\St=20$,
  showing the existence of regions where particles have different
  velocities (highlighted by gray and black arrows).}
\end{figure}

\noindent Two effects have recently been singled out to explain the
speed-up of collisions between small finite-size particles suspended
in turbulent flows~\cite{sc97,s03,ffs02}: \emph{preferential
  concentration}, that is the development of strong inhomogeneities in
their spatial distribution
(Fig.~\ref{fig:twomechanisms}a)~\cite{wwz98,rc00,gv08}, and the
formation of \emph{fold caustics} (also called the \emph{sling
  effect}), which results in large probabilities that close particles
have important velocity differences
(Fig.~\ref{fig:twomechanisms}b)~\cite{bccm05,wm05-wmb06,fp07}. Improving
the collision kernels used in kinetic models for atmospheric physics,
astrophysics, and engineering requires quantifying precisely the
individual contribution of these two effects and, in particular, to
what extent turbulence might affect them, i.e. how they depend on the
Taylor-scale Reynolds number $\Rey_\lambda$ of the
flow~\cite{dft08,xwg08}.

In this Letter, to straighten out these questions, we consider
suspensions of small, heavy, and dilute particles, which is a setting
relevant to the early stage of rain droplets formation in
clouds~\cite{ffs02}. In these conditions, particles are simply dragged
by viscous forces and each individual trajectory $\bm X(t)$ solves the
equation
\begin{equation}
  \tau \ddot{\bm X} = -\dot{\bm X} + \bm u (\bm X,t)\,,
  \label{eq:partdynamics}
\end{equation}
where dots denote time derivatives, $\bm u$ the fluid velocity field,
solution of the incompressible Navier--Stokes equation, and $\tau =
2a^2\alpha/(9\nu)$ the Stokes time, depending on particle radius $a$,
density contrast $\alpha$ with the fluid, and fluid kinematic
viscosity $\nu$, see \cite{tb09} for a recent review. The importance
of inertia in the particle dynamics is quantified by the \emph{Stokes
  number} $\St = \tau/\tau_\eta$, where $\tau_\eta$ is the fluid eddy
turnover time associated to the Kolmogorov dissipative scale $\eta$.
The collision rate between particles is evaluated using the
ghost-particle approach~\cite{rc00}, which assumes that particles can
occupy any point of space independently of the positions of
others. This approximation is valid in the asymptotics of very diluted
suspensions, and has the advantage of relying on stationary dynamical
statistics: the geometrical collision rate is then determined by the
value at $r=2a$ of the \emph{approaching rate}~\cite{bccm05}
\begin{equation}
  \kappa(r ; \St) = -\left\langle \left. \dot{R} \ \right | \, R\!=\!
  r \mbox{ and } \dot{R}\le 0 \right\rangle \, p_2(r)\,.
  \label{eq:defkappa}
\end{equation}
Here $R$ denotes the distance between two particles with Stokes number
$\St$, and $p_2$ its probability density. The average is performed
over time and particle pairs, with the condition to be separated by a
distance $r$ and to approach each other.  Clearly the behavior of
$\kappa(r;\St)$ prescribes the dependence of the collision rate upon
the particle attributes (size and mass density contrast). Caustics and
preferential concentration (Fig.~\ref{fig:twomechanisms}) intricately
appear in~(\ref{eq:defkappa}) affecting the conditional average of the
velocity difference $\dot{R}$ and the $r$-dependence of $p_2$,
respectively.  In particular, as shown in~\cite{bbbclmt07}, $p_2(r)$,
which is straightforwardly related to the radial distribution function
of~\cite{rc00}, behaves as a power law in the dissipative range,
namely $p_2(r) \propto r^{\mathrm{D}_2-1}$ for $r\ll\eta$, where
$\mathrm{D}_2\in [0\!:\!3]$ is the \emph{correlation dimension} of the
particle distribution and non-trivially depends on the Stokes number.

We focus here on the velocity contribution by studying the behavior as
a function of the separation $r$ of the longitudinal particle velocity
structure functions
\begin{equation}
  S_p(r;\St) = \left\langle \left. |\dot{R}|^p \ \right | \, R\!=\!  r
  \right\rangle.
  \label{eq:defstructfn}
\end{equation}
The choice of defining structure functions with absolute values is
motivated by the definition of the collision kernel (via the
approaching rate), since we do not expect important differences
between negative and positive velocity fluctuations.  One can
therefore estimate: $\kappa(r) \propto r^{\mathrm{D}_2-1} S_1(r;\St)$
(see \cite{bccm05}). Evaluating $S_p(r;\St)$ for values of $p$
different from $1$, besides providing a more complete characterization
of the velocity statistics, allows one to account also for
fluctuations of the local approaching rate, which can vary
significantly from place to place.  In the limit of small inertia, the
particle dynamics approaches that of tracers and consequently the
velocity difference $\dot{R}$ is essentially coincident with the fluid
longitudinal increment over a separation $R$. Conversely, when
$\St\to\infty$, particles move ballistically in the flow with
uncorrelated velocities and the structure functions $S_p(r;\St)$
become independent of $r$. For intermediate values of the Stokes
number, one expects a non-trivial behavior of $S_p$ as a function of
$r$ and $\St$. Data analyzed in this study are from a direct numerical
simulation at $\Rey_\lambda\approx 400$ described in
\cite{bblst09,icfd}.

\begin{figure}[b!]
  \vspace{-15pt}
  \includegraphics[width=\columnwidth]{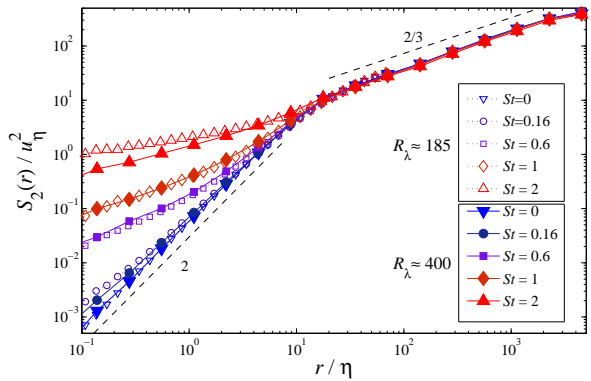}
  \vspace{-25pt}
  \caption{\label{fig:compareS2_2Re} (color online) Second-order
    longitudinal velocity structure function for particles with
    various Stokes numbers and for two Reynolds numbers.}
\end{figure}
Figure~\ref{fig:compareS2_2Re} represents the behavior of the
second-order structure function $S_2(r;\St)$, measured in direct
numerical simulations, for two different values of the carrier flow
Reynolds number (see~\cite{bblst09} for details on the
simulations). One distinguishes different regimes, depending whether
$r$ is within the dissipative or inertial range of the turbulent
carrier flow. While the dissipative-range behavior directly relates to
inter-particle collisions, the inertial-range behavior has important
implications on the relative dispersion of particles in turbulent
flow~\cite{bblst09} in general and for pollution models in
particular. In the sequel we investigate these two regimes separately.

In the dissipative range, the structure functions display a power-law
behavior $S_p(r;\St) \propto r^{\xi_p}$. The two asymptotics of weak
and strong inertia imply that $\xi_p\simeq p$ for $\St\ll1$ and
$\xi_p\to0$ for $\St\to\infty$. For intermediate values of the Stokes
number, $p\mapsto\xi_p(\St)$ is a convex function of the order $p$
with values in $[0:p]$.
\begin{figure}[t!]
  \vspace{-10pt}
  \includegraphics[width=\columnwidth]{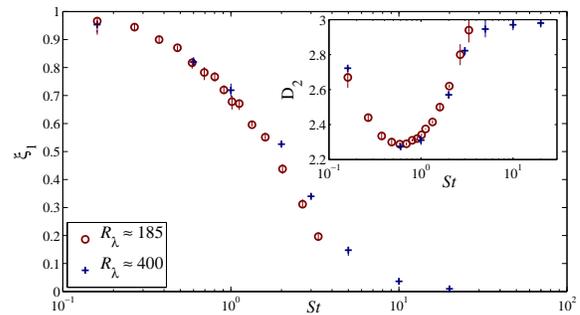}
  \vspace{-25pt}
  \caption{\label{fig:exponents2Re} (color online) Scaling exponent,
    $\xi_1$, vs.\ $\St$ for two values of $\Rey_\lambda$. Inset:
    correlation dimension $\mathrm{D}_2$ vs.\ $\St$.}
\end{figure}
Figure~\ref{fig:exponents2Re} shows the first-order exponent $\xi_1$
as a function of the Stokes number. One can clearly observe that for
$\St=O(1)$, the exponent $\xi_1$ takes non-trivial values spanning the
whole interval $[0\!:\!1]$. The present accuracy of data does not
allow for settling either the issue of a possible saturation of the
exponent to the limiting values at the two extrema, nor a possible
dependence of the exponent upon $\Rey_\lambda$. Despite a factor two
in $\Rey_\lambda$, data differ by less than the errors made in the
determination of the exponents or in the value of $\tau_\eta$ that
enters the definition of $\St$.

At first glance the continuous variation of the exponent $\xi_1$ from
1 to~0 at increasing $\St$ seems inconsistent with a naive picture of
the role of caustics in velocity statistics. Fold caustics are a part
of catastrophe theory~\cite{asz82}; they occur when fast particles
catch up with slower ones to create regions where several velocities
can be found at the same location as in
Fig.~\ref{fig:twomechanisms}b. If particles conserve their velocity
and move ballistically, such caustics will extend over the whole
domain (whence the analogy with caustics formed by light
rays~\cite{wm05-wmb06}). The typical velocity difference between two
particles becomes in that case independent of their distance, meaning
that structure functions tend to a constant as $r\to0$, and thus
$\xi_p=0$. However, there are two clear reasons why this
continuous-field picture may fail. First, because of their dissipative
dynamics, particles concentrate on dynamical attractors in the
position-velocity phase space~\cite{bccm05}. Such sets are fractal and
correlated with the fluid and lead to the formation of caustics of
various strength with non-trivial probabilities. Second, as the
particle velocity relaxes to the fluid velocity, the spatio-temporal
extent of such caustics may also have complex statistical properties.

\begin{figure}[t!]
  \vspace{-15pt}
  \includegraphics[width=1.\columnwidth]{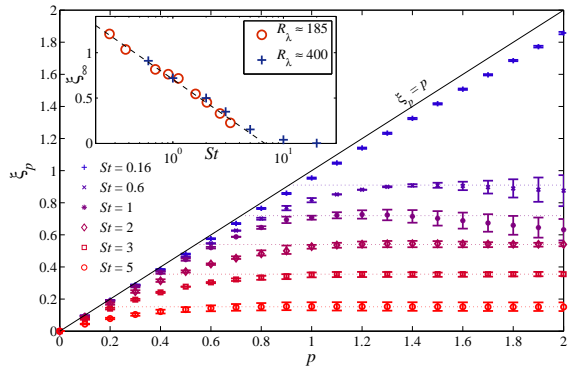}
  \vspace{-25pt}
  \caption{\label{fig:exponents_xip} (color online) Scaling exponents
    $\xi_p$ of the particle velocity structure functions $S_p$ for
    various $\St$ and $\Rey_\lambda\approx400$. Inset: saturation
    exponents $\xi_\infty$ as a function of $\St$ for two values of
    $\Rey_\lambda$. Exponents are obtained by measuring the mean
    logarithmic derivative of $S_p(r)$ in $0.2\!\le\!
    (r/\eta)\!\le\!2$; errors correspond to the largest deviations
    observed in the fitting range.}
  \vspace{-20pt}
\end{figure}

To better quantify the contribution from caustics, we extend our
investigation to the particle velocity scaling exponents $\xi_p$'s
with orders $p$ other than 1, shown in Fig.~\ref{fig:exponents_xip}
for various values of the Stokes number.  At small orders, the
exponents are almost tangent to the line $\xi_p=p$ while, at large
orders, they approach or saturate to an asymptotic value $\xi_\infty$,
monotonically decreasing with $\St$ as shown in the inset. Numerical
data suggests that $\xi_\infty \propto \ln (1/\St)$ for $\St\lesssim
7$ and that $\xi_\infty \simeq 0$ for $\St\gtrsim 7$. The current
numerical accuracy does not enable distinguishing between a sharp and
smooth transition at $\St\simeq 7$.
However, we notice that the estimated value of such a critical Stokes
number is very close to that for which $\mathrm{D}_2\approx 3$ (see
inset of Fig.~\ref{fig:exponents2Re}). As discussed in \cite{dft08} a
saturation of $\mathrm{D}_2$ to the space dimension is indeed expected
for $\St$ values at which caustics become dominating.  In this
respect, we also notice that the saturation exponent $\xi_\infty$ can
be interpreted as the co-dimension of large fold caustics associated
to order-unity velocity jumps. Indeed, such caustics contribute to the
structure function $S_p(r;\St)$ a term of the form $V_p\,P(r)$ where
$V_p$ is the $p$-th order moment of the velocity difference inside the
caustics and $P(r)$ is the probability of having such a caustic
present in a box of size $r$. The saturation of the scaling exponents
suggests that $P(r)\propto r^{\xi_\infty}$, so that
$\mathrm{D}^{\mathrm{(c)}} \equiv 3-\xi_\infty$ is the (statistical)
Hausdorff dimension of the set of caustics. At smaller orders, the
statistics is dominated by other events for which one can figure out
two conceivable scenarios. A first possibility is that caustics
distribution spans all possible sizes with non-trivial co-dimensions,
i.e.\ is a multi-fractal. In this case they affect all orders and give
rise to multiscaling and to a non-trivial behavior of the exponents
$\xi_p$ before the saturation ~\cite{clmv00}. The alternative
possibility is that the caustics are randomly distributed with a
typical size and dominate the velocity statistics at large moments
only, while small orders are controlled by the smooth regions of the
particle velocity. In that case the structure function would display a
bi-fractal behavior similar to that present in random solutions to the
Burgers equation (see, e.g., \cite{bk07}), namely $\xi_p = p$ for
$p\le \xi_\infty$ and $\xi_p = \xi_\infty$ for $p\ge
\xi_\infty$. Current numerical results do not permit to distinguish
between these two possibilities. As seen from
Fig.~\ref{fig:exponents_xip}, the measured exponents show some
deviations from the bi-fractal behavior.  However as already observed
in other settings~\cite{bclst04-mbpf05}, this apparent multiscaling
could be an artifact due to the presence of sub-leading terms or
logarithmic corrections.

To further disentangle the question of the contribution of caustics to
velocity scaling, we investigate the statistical properties of $\sigma
= \dot{R}/R$, which can interpreted as a longitudinal velocity
gradient of an effective particle velocity field.  This quantity is at
the center of much work devoted to the relative motion of a pair of
particles in time-uncorrelated flows
\cite{wm05-wmb06,p02,dftt07,bcht08}. There, the dynamics of $\sigma$
becomes independent of $R$ at very small scales, a far from obvious
feature for particles transported by real flows, where time
correlations and structures play important roles.  Further, results in
random flows suggest that the conditional probability density
$p(\sigma\,|\,R\!=\!r)$ is independent of $\sigma$ at small scales and
has power-law tails.
\begin{figure}[t!]
  \vspace{-15pt}
  \includegraphics[width=\columnwidth]{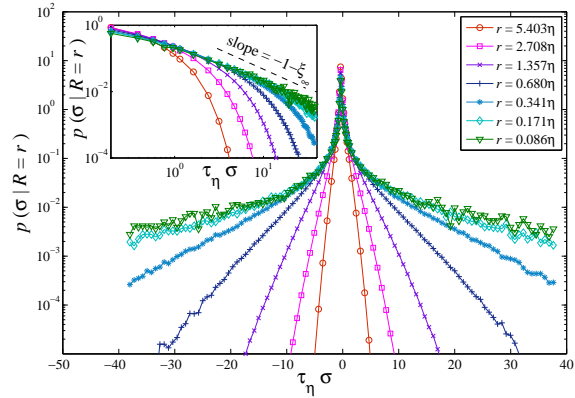}
  \vspace{-25pt}
  \caption{\label{fig:pdfsig_st15} (color online) Probability density
    of the (rescaled) longitudinal velocity difference $\sigma =
    \dot{R}/R$ for various values of $r$ and for $\St = 3.3$ and
    $\Rey_\lambda\approx 185$. Inset: same for the right tail in
    log-log coordinates.}
  \vspace{-20pt}
\end{figure}
As seen from Fig.~\ref{fig:pdfsig_st15}, numerical measurements in
turbulent flows suggest features similar to those of structure-less
random flows. The core of the distributions associated to different
scales $r$ collapse for $|\sigma|\lesssim\sigma^\star(r)$ on a
distribution with a fat, almost algebraic behavior. Interestingly, the
associated power-law exponent is close to $-(1+\xi_\infty)$,
suggesting that $\langle \sigma^p \rangle$ diverges for
$p>\xi_\infty$, a behavior favoring the bi-fractal scenario. Indeed
$S_p(r;\St) = r^p \langle \sigma^p \,|\, R \!=\! r \rangle \simeq r^p
\langle \sigma^p \rangle$ for $r\to0$ and $p$ such that $\langle
\sigma^p \rangle < \infty$. However, for $|\sigma|\gtrsim
\sigma^\star(r)$, the distributions display stretched-exponential
tails, whose contribution to the structure function is for the moment
unsettled. They are connected to the caustics size probability
distribution and could lead to multiscaling. A related open question
is the non-trivial entanglement between clusters and large velocity
differences, as already stressed in random flows~\cite{o08}. This
latter feature might imply an intricate dependence on $r$ of velocity
difference statistics, that might lead to multiscaling. Settling
numerically the issue of bi- versus multiscaling would require to
explore systems were statistics can be handled in a more systematic
way, as for instance in random correlated flows or real flows at
smaller Reynolds numbers.

\begin{figure}[t!]
  \includegraphics[width=\columnwidth]{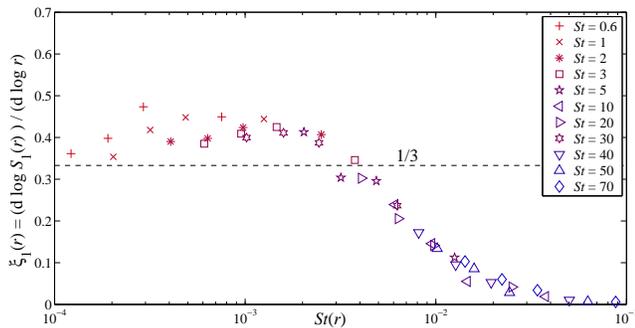}
  \vspace{-25pt}
  \caption{\label{fig:localexponent}(color online) First-order local
    exponent $\xi_1(r)$ as a function of the local Stokes number
    $\St(r)$ for $\Rey_\lambda \approx 400$. The horizontal dashed
    line represents tracers K41 expectation.}
  \vspace{-5pt}
\end{figure}
We finally turn to the behavior of the velocity structure function for
separations within the inertial range of turbulence, i.e.\ for
$\eta\ll r\ll L$. As seen from Fig.~\ref{fig:compareS2_2Re}, particle
velocity structure functions recover the fluid ones when $r$ becomes
very large. Indeed as $r$ increases the associated eddy turnover time
grows as $r^{2/3}$ (where we used the Kolmogorov 1941 scaling) so that
the effective strength of inertia reduces. Similarly to random
self-similar carrier flows~\cite{bcht08}, this effect can be put on a
quantitative ground in terms of a scale-dependent Stokes number
$\St(r) = \varepsilon^{1/3}\tau/r^{2/3}$ defined as the ratio between
the particle response time and the turnover time associated to the
scale $r$, where $\varepsilon$ denotes the mean dissipation rate of
kinetic energy. We check whether the local scaling exponent
$\xi_p(r;\tau) \equiv (\mathrm{d}\ln S_p(r;\St))/(\mathrm{d}\ln r)$
does depend on $\St(r)$ only, as observed in random self-similar flows
\cite{bcht08}. Figure~\ref{fig:localexponent} shows a good collapse of
the values of $\xi_1(r,\tau)$ associated to various $\tau$ and of $r$,
once represented as a function of $\St(r)$. Moreover, the curve
$\xi_1(\St(r))$ has a shape qualitatively very similar to that of
$\xi_1(\St)$ observed in the dissipative range and shown in
Fig.~\ref{fig:exponents2Re}, this fact is relevant to heavy particle
dispersion in turbulent flows~\cite{bblst09}. Let us stress that data
corresponding to small $\St(r)$ in Fig.~\ref{fig:localexponent} show
deviations from the K41 scaling that are similar to those expected for
tracers-like statistcs.

To conclude we briefly discuss the applicability of the present
results to atmospheric physics.  The main shortcoming of the proposed
approach is that the gravity force is neglected.  As observed
in~\cite{arw08} for dynamics of water droplets in warm clouds,
gravitational settling is found to dominate the statistics of velocity
differences between particles. However, this effect acts mainly
between particles of different sizes that fall at different
speeds. Present results should apply to earlier stage of rain
formation during which the droplets are almost mono-disperse and might
play an important role in explaining the observed fast spectral
broadening observed in clouds.

This study benefited from fruitful discussions with L.~Collins,
G.\ Falkovich, B.\ Mehlig, L.-P.\ Wang, and M.\ Wilkinson. JB and AL
acknowledge support from NSF under grant No.\ PHY05\_51164. Part of
this work was supported by ANR under grant
No.\ BLAN07-1\_192604. Simulations were performed at CASPUR and CINECA
(Italy), and in the framework of the DEISA Extreme Computing
Initiative supported by the DEISA Consortium (co-funded by the EU, FP6
project 508830).

\end{document}